\documentclass[%
 reprint,
 amsmath,amssymb,
 aps,
]{revtex4-2}

\usepackage{graphicx}
\usepackage{dcolumn}
\usepackage{bm}
\usepackage{graphicx}

\usepackage{dsfont}

\begin{document}

\title{Wave-Particle Duality in the Negative Information Sea}

\author{Daegene Song}
 \affiliation{Department of Management Information Systems, Chungbuk National University, Cheongju, 28644 Korea.}

\date{\today}

\begin{abstract}

Quantum theory reveals astonishing and counterintuitive phenomena not found in classical physics, such as wave-particle duality, where entities like electrons and photons exhibit both wave-like and particle-like behaviors. In this paper, we leverage advancements in quantum information science to gain new insights into this phenomenon. We specifically examine negative conditional entropy in quantum entanglement, where the selection of the measurement basis appears to ripple backward in time, akin to Dirac's model of an infinite sea of negative energy states filled with electrons, where holes in this sea appear as positrons. We propose that an observer's knowledge of the measurement choice, analogous to a hole in the negative information sea, corresponds to the wave aspect of the system, while the classical outcome aligns with its particle nature. This exploration of the relationship between microscopic quantum phenomena and macroscopic observations offers new perspectives on the mind-matter duality.

\end{abstract}

\maketitle

\section{Introduction}

Wave-particle duality, a cornerstone of quantum theory, asserts that quantum entities like electrons and photons simultaneously exhibit properties of both waves and particles \cite{young,taylor,berthold,marlan}. This phenomenon is vividly demonstrated in experiments such as the double-slit experiment \cite{steeds,nairz,chen,zeilinger}, where particles create an interference pattern indicative of wave behavior when undetected, but act as discrete particles when measured. This dual nature challenges classical physics, which treats waves and particles as distinct entities, and has spurred extensive research \cite{yoon,angelo,coles,qian,zela,qian2,xue} to understand the principles governing this phenomenon and its implications for the nature of reality.

Recent breakthroughs in quantum information science \cite{deutsch,QC1,QC2} have paved new pathways for delving into the core principles of quantum mechanics. Developments in quantum computing and technology, particularly the concepts of quantum superposition and entanglement, reveal how quantum systems can process information in ways that classical systems cannot \cite{shor,bb84}. These advancements, including quantum teleportation \cite{tele,bouw}, superdense coding \cite{super,schaetz,williams}, and quantum error correction \cite{error}, exploit quantum mechanics' unique properties to achieve unprecedented feats in communication security \cite{comm} and computational power \cite{ladd,preskill}.

\begin{figure}
\begin{center}
\includegraphics[width=.4\textwidth]{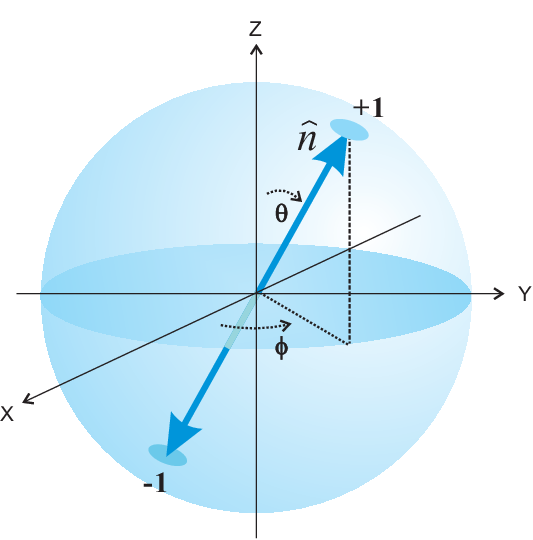}

\end{center}
\caption{ A choice of observable for measuring a given qubit can be envisioned as a vector \(\hat{n}\) pointing in the direction \((\theta, \phi)\) on the Bloch sphere, representing the observer's choice of reference frame.}
\label{Bloch}
\end{figure}

Can we deepen our understanding of wave-particle duality through the lens of advancements in quantum information? This paper posits that we can. A key tool for quantifying the relationship between two random variables is conditional entropy \cite{slepian}, which measures the uncertainty of a variable \( B \) given the knowledge of another variable \( A \). This represents the average additional information needed to describe \( B \) once \( A \) is known. Remarkably, quantum information theory introduces the notion of negative conditional entropy \cite{cerf, horodecki, vedral}, a concept absent in classical physics. This negativity is linked to quantum entanglement, where one subsystem can contain more information about the other subsystem than classical physics allows. By examining the idea of negative entropy, this paper aims to uncover new insights into wave-particle duality, providing a fresh perspective on this intricate aspect of quantum mechanics.

\section{Quantum Information}

Wave-particle duality in quantum information theory is manifested through the principle of superposition, where a qubit, the fundamental unit of quantum information, simultaneously exists in a combination of both its basis states. With Pauli matrices \( \vec{\sigma} = (\sigma_x, \sigma_y, \sigma_z) \), a qubit, represented as \( |\psi\rangle\langle\psi| = \frac{1}{2}(\mathds{1} + \hat{m} \cdot \vec{\sigma}) \), can be visualized as a unit vector \( \hat{m} = (\sin\vartheta\cos\varphi, \sin\vartheta\sin\varphi, \cos\vartheta) \)  pointing in the \( (\vartheta, \varphi) \) direction on the Bloch sphere, with \( 0 \leq \vartheta \leq \pi \) and \( 0 \leq \varphi < 2\pi \). The wave-like behavior is evident in the interference patterns and entanglement that arise from qubits being in superposition, reflecting their ability to encode and process multiple possibilities at once \cite{deutsch,jozsa}.

An observable in quantum mechanics is a physical quantity that can be measured, such as position, momentum, or spin. Observables are associated with Hermitian (self-adjoint) operators, and the eigenvalues of these operators correspond to possible measurement outcomes. In the case of qubit measurement, an observable may be written as follows:
\begin{equation}
\hat{\mathcal{O}}_{\hat{n}} \equiv |\hat{n}\rangle\langle\hat{n}| - |\hat{n}^{\perp}\rangle\langle\hat{n}^{\perp}|
\label{obser}
\end{equation}
where \( |\hat{n}\rangle = \cos \frac{\theta}{2}|0\rangle + e^{i\phi}\sin\frac{\theta}{2}|1\rangle \) and \( |\hat{n}^{\perp}\rangle \) is the state orthogonal to \( |\hat{n}\rangle \). In analogy with the qubit notation described above, the choice of the observable in (\ref{obser}) may be considered equivalent to choosing a unit vector \( \hat{n} \) pointing in the \( (\theta, \phi) \) direction on the Bloch sphere (Figure \ref{Bloch}) as a measurement basis, where \( 0 \leq \theta \leq \pi \) and \( 0 \leq \phi < 2\pi \). Therefore, in the context of measuring a qubit, \( \hat{n} \) corresponds to a reference frame for the observer \cite{song} (also see \cite{fuchs}).

Von Neumann's idealized measurement framework, as described in his foundational work \cite{neumann}, addresses the process of measurement in quantum mechanics. A qubit and a measuring apparatus, both in their initial states, can be described as follows:
\begin{equation}
|\psi_0\rangle_Q \otimes |m_0\rangle_M
\end{equation}
If we assume that an observer chooses the \( \hat{n} \) direction to measure the system \( Q \), the measurement process can be modeled through an interaction Hamiltonian that couples the quantum system and the measurement apparatus: \( \hat{H} = \hat{\mathcal{O}}_{\hat{n}} \cdot \hat{P} \), where the observable \( \hat{\mathcal{O}}_{\hat{n}} \) is given in (\ref{obser}) and \( \hat{P} \) is the momentum operator that changes the macroscopic apparatus pointer. This interaction causes the state of the apparatus to become correlated with the state of the quantum system as follows:
\begin{equation}
U: |\psi_0\rangle_Q |m_0\rangle_M \rightarrow c_0 |\hat{n}\rangle_Q |m_1\rangle_M + c_1 |\hat{n}^{\perp}\rangle_Q |m_2\rangle_M
\label{corr}
\end{equation}
where \( |m_1\rangle \) and \( |m_2\rangle \) are classically distinguishable pointers of the apparatus, appearing with probabilities \( |c_0|^2 = |\langle \hat{n}|\psi_0\rangle |^2 \) and \( |c_1|^2 = |\langle \hat{n}^{\perp}|\psi_0\rangle |^2 \), respectively.

Note that through the interaction between the system and the apparatus, as seen in (\ref{corr}), the observer's measurement choice \(\hat{n}\) is now encoded in \(Q\). Interestingly, while \(\hat{n}\) has an infinite number of possible values, defined by \(\theta\) and \(\phi\) on the sphere (Figure \ref{Bloch}), the outcome of the apparatus is either up or down, or \(\pm 1\). Thus, after the interaction in (\ref{corr}), the continuity in the system \(Q\) is correlated with a finite discrete value in the apparatus \(M\). This seemingly peculiar interconnection will be important for our subsequent discussion of wave-particle duality.

\section{Conditional Entropy and Negative Information Sea}

Information theory is a mathematical framework designed to quantify the transmission, processing, and storage of information \cite{shannon,ash}. It was developed to understand and optimize communication systems. A key concept in this field is conditional entropy, a measure of the amount of uncertainty or information remaining about one part of a system when the state of another part is known.  It's a way of quantifying the interdependence between different parts of a system and how information about one part reduces our uncertainty about another part.

The quantum analog of conditional entropy for a bipartite quantum state of \(A\) and \(B\) is defined using the von Neumann entropy: \(S(A|B) = S(AB) - S(B)\). Quantum conditional entropy measures the amount of uncertainty remaining about one quantum system given knowledge of another. Surprisingly, unlike its classical counterpart, quantum conditional entropy can be negative \cite{cerf,horodecki}: for instance, in the case of a Bell state, \( \frac{1}{\sqrt{2}} \left( |00\rangle + |11\rangle \right)_{AB} \), the conditional entropy can be calculated as \( S(A|B) = 0 - 1 = -1 \). Negative entropy suggests that knowing more about one part of a quantum system can paradoxically increase the overall uncertainty \cite{vedral}.


\begin{figure}
\begin{center}
\includegraphics[width=.4\textwidth]{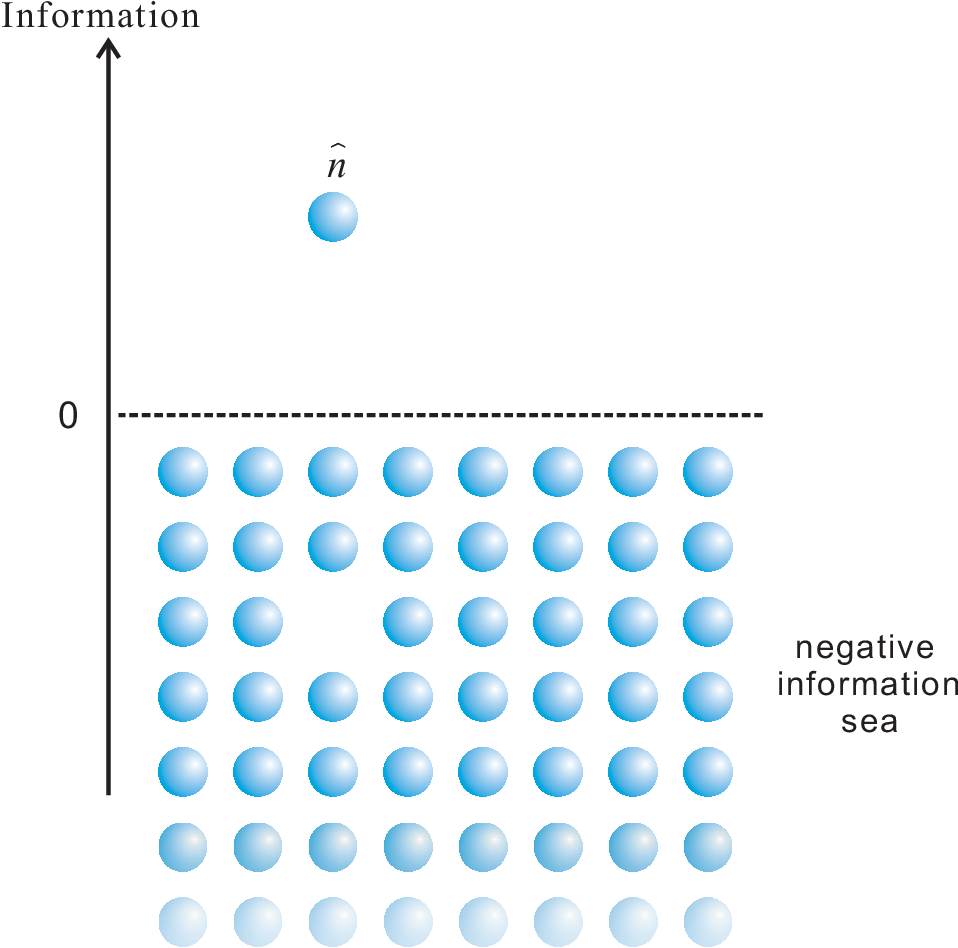}

\end{center}
\caption{ Analogous to Dirac's negative energy sea, the emergence of negativity in quantum conditional entropy can be envisioned as a hole in the negative information sea. Similar to the prediction of the positron, this hole corresponds to the observer's knowledge of the basis choice $\hat{n}$ when observing the apparatus outcome. }
\label{Sea}
\end{figure}

A historical parallel to this odd phenomenon can be found in the early development of quantum mechanics. In the late 1920s, Paul Dirac was working on a relativistic wave equation for electrons that produced negative energy solutions. Rather than dismissing these results as unphysical, Dirac introduced the concept of what was later called the Dirac sea—a vacuum filled with all possible negative energy states \cite{dirac}. The Pauli exclusion principle prevents electrons from decaying into these negative energy states. Dirac's theory implied the existence of \lq holes\rq\, in this sea, behaving like particles with positive energy but opposite charge, later identified as positrons and confirmed experimentally.

Similarly, the concept of an antiqubit \cite{cerf} proposes a quantum state that, much like antiparticles annihilating particles, could neutralize or cancel out information when combined with a corresponding qubit. This idea aligns with the phenomenon of negative conditional entropy in quantum systems, where entanglement between subsystems reduces uncertainty beyond classical limits. Essentially, one part of the system carries negative information about the other. An antiqubit is thus envisioned as an elusive quantum state embodying this negative information, reducing the total uncertainty when interacting with a qubit.

\begin{figure}
\begin{center}
\includegraphics[width=.4\textwidth]{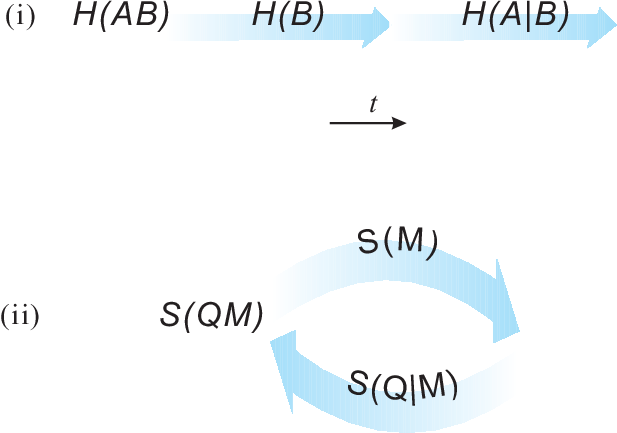}

\end{center}
\caption{ (i) In the classical case, measuring system $B$ reduces the total uncertainty about $A$ and $B$, leading to the uncertainty about $A$ given $B$. (ii) Unlike the classical case, quantum conditional entropy can take negative values when $Q$ and $M$ are entangled, as seen in the measurement process in (\ref{corr}). This suggests that observing $M$ could be followed by a time-reversal process of the conditional entropy $S(Q|M)$. }
\label{Nega}
\end{figure}

Following Dirac's approach, we propose that the vacuum is filled with negative information (Figure \ref{Sea}). Moreover, similar to how a hole in the Dirac sea corresponds to an antiparticle, a hole in the negative information sea is suggested to be an antiqubit. In particular, in the context of the quantum measurement process involving the correlation of the system and the apparatus described in (\ref{corr}), the conditional entropy \(S(Q|M)\) represents the negative information carried by the qubit \(Q\) with respect to the outcome of \(M\). This implies that the quantum system encoding the observer's basis choice \(\hat{n}\) travels backward in time as an antiqubit (Figure \ref{Nega}).

Classical wave theories require a medium, such as water for ocean waves or air for sound waves, to propagate. In contrast, quantum waves do not need a physical medium; they describe probabilities and amplitudes of states. The medium in quantum theory, i.e., the negative information sea as proposed in this paper, can be interpreted as follows: quantum waves are represented by wavefunctions existing in an abstract Hilbert space rather than physical space. Thus, it is reasonable to view this {\it{imaginary}} space as corresponding to the observer's mind, rather than the three-dimensional classical world where the outcome of the apparatus is experienced.

As far as qubit measurement is concerned, the hole in the negative information sea corresponds to an antiqubit in observing the classical outcome of the apparatus.  That is, the quantum system \(Q\) in (\ref{corr}) corresponds to the observer's knowledge of \(\hat{n}\), or a wave-like aspect, serving as a context for observing the physical outcome of \(M\), i.e., \(\pm 1\), a particle-like element. Interestingly, in quantum theory, the Heisenberg cut delineates the boundary between the microscopic realm, where quantum effects like superposition and entanglement are prevalent, and the macroscopic realm, where classical physics dominates. The analogy presented in this paper offers a perspective on wave-particle duality as a duality between mind (i.e., microscopic domain) and matter (i.e., macroscopic realm).

\section{Remarks}

Plato's theory of ideal forms, also known as \lq Forms\rq\,  or \lq Ideas,\rq\,  suggests that beyond the physical world we perceive with our senses, there exists a non-material realm of perfect, unchanging forms that represent the true essence of all things. According to Plato, these Forms are the ultimate reality, while the objects we encounter in the physical world are merely imperfect copies of these ideal forms. While we cannot empirically verify the existence of this ultimate reality, we can conceive of these ideal forms in our minds, as we are capable of theoretically manipulating perfect entities like circles and squares.

Similarly, mind-matter duality explores the relationship between the mental and physical realms, questioning how consciousness (the mind) and the material world (matter) interact and coexist. This duality suggests that mind and matter are fundamentally distinct yet interconnected. In Cartesian dualism, famously proposed by Ren${\acute{\rm{e}}}$ Descartes, the mind is seen as a non-physical substance that can influence the physical body, and vice versa. This view raises questions about how two different substances can causally interact.

In a different domain, wave-particle duality, a cornerstone of quantum mechanics, reveals the counterintuitive nature of quantum entities. This duality challenges classical intuitions and underpins the field of quantum information science, which explores how quantum systems can process information in ways that classical systems cannot. This paper discussed how negative entropy arose from the entanglement between the microscopic system and the macroscopic apparatus. Specifically, considering the system that encoded the observer's choice of measurement basis evolving backward in time provided the context for observing the classical outcome associated with the apparatus.

\end{document}